\documentclass{webofc}
\usepackage[varg]{txfonts}   
\begin{document}
\title{Measurement of charged and full jet production\\and nuclear modification factor in pp and p--Pb collisions\\with ALICE}

\author{
     \firstname{Austin} \lastname{Schmier for the ALICE collaboration}\inst{1}\fnsep\thanks{\email{a.schmier@cern.ch}}
}

\institute{
     University of Tennessee, Knoxville
}

\abstract{
     The study of jet production in small collision systems is essential for testing our understanding of perturbative and non perturbative QCD and cold nuclear matter (CNM) effects. In addition, studies at high multiplicity in small collision systems exhibit signatures of collectivity, which is still not fully understood within a unified picture across system size. Jet quenching in small systems is not observed within current measurement precision, calling for more precise jet measurements. Results of charged particle and fully reconstructed jet production in pp and p--Pb collisions are presented at $\sqrt{s_{\rm NN}} = 5.02$ and $8$ TeV, along with the corresponding nuclear modification factor $R_{\rm pPb}$ at $\sqrt{s_{\rm NN}} = 5.02$ TeV. These results are the most precise measurements of the $R_{\rm pPb}$ by ALICE to date. To investigate whether jet energy is redistributed in CNM, the cross-section ratios for different jet resolution parameters ($R$) are compared between pp and p--Pb collisions, as well as within each collision system. Finally, comparisons between data and model predictions are discussed. This result extends previous LHC measurements to lower jet transverse momentum, constraining hard parton production and fragmentation mechanisms applied in model calculations, and the impact of the nuclear-modified parton distribution function on jet production. This measurement also provides new constraints on jet quenching in small collision systems.  
}

\maketitle

\section{Introduction}
\label{introduction}

In high-energy collisions, quarks and gluons, governed by quantum chromodynamics (QCD), can scatter with a large momentum transfer, forming sprays of particles called jets. Jet production offers insights into perturbative (pQCD) and non-perturbative QCD, serving as a crucial probe for studying the properties of cold nuclear matter and quark-gluon plasma (QGP) in heavy-ion collisions, a state of matter where partons are no longer confined into hadrons. This research is vital for understanding the evolution of the early universe. The focus of these proceedings is on studying jet production in small collision systems, traditionally not expected to form QGP, though recent studies indicate evidence of collectivity\cite{ALICE:2023ama}.

ALICE (A Large Ion Collision Experiment) at CERN LHC, a general-purpose detector specializing in heavy ions, employs two methods for jet reconstruction: charged particle jets (charged jets), utilizing charged tracks from the Time Projection Chamber (TPC) and Inner Tracking System (ITS), and fully reconstructed jets (full jets), incorporating information from both charged tracks and neutral particles detected by the Electromagnetic Calorimeter (EMCal). Additionally, the EMCal acts as a trigger detector for jet events.

Charged jets benefit from full azimuthal coverage and broader pseudorapidity acceptance than full jets, benefiting from precise tracking for tighter model constraints. However, charged jets are limited by the momentum reach of the TPC.  In contrast, full jets, reconstructed with EMCal triggers, achieve higher momentum reach, but face constraints due to limited EMCal acceptance in azimuth and smaller pseudorapidity coverage. Full jets align more closely with theoretical definitions but are subject to biases introduced by EMCal triggers, necessitating corrections and adding uncertainties.

\section{Cross-Sections, Ratios, and the Nuclear Modification Factor}
\label{body}

Jet formation involves initial conditions of the partons, hard scattering of the partons, fragmentation, and subsequent hadronization. The parton distribution function (PDF) defines initial conditions, indicating the probability of finding a parton with a specific momentum fraction in a nucleon. Hard scattering is governed by perturbative QCD (pQCD). Fragmentation is characterized by fragmentation functions, representing the likelihood of a parton fragmenting into a hadron with a particular momentum fraction. Hadronization, a non-perturbative process, is also described by QCD. The cross-section, sensitive to all these stages, is valuable for constraining jet production models.

\begin{figure*}
     \centering
     \includegraphics[width=0.98\textwidth]{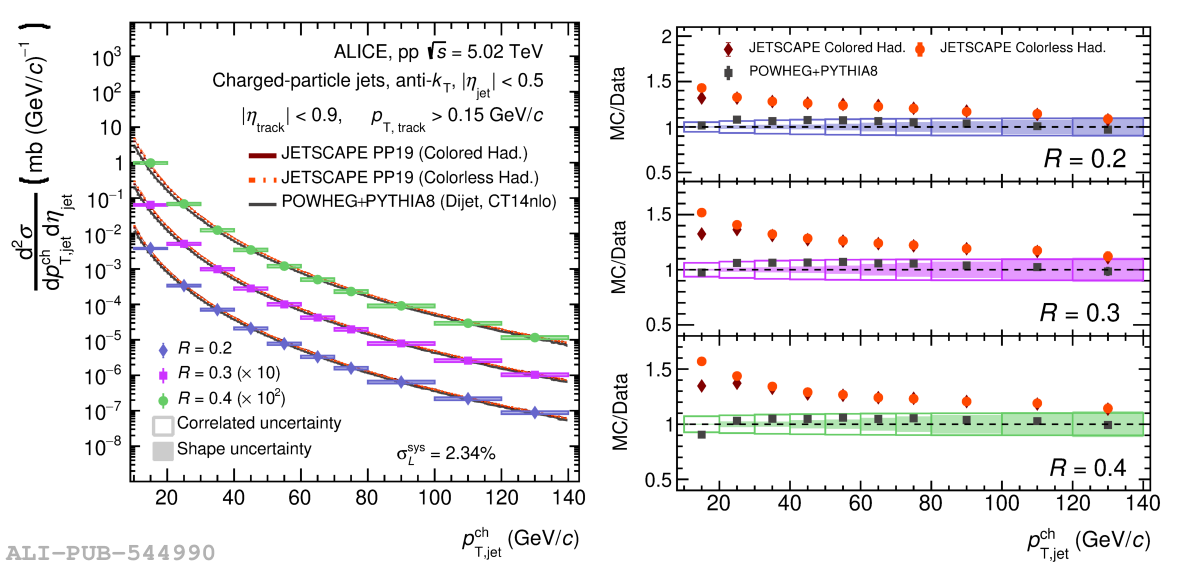}
     \caption{Charged particle jet cross-sections for R = 0.2 - 0.4 at $\sqrt{\it{s}}$ = 5.02 TeV in pp collisions from ALICE compared to POWHEG+PYTHIA8 and two JETSCAPE calculations, scaled by an arbitrary integer for visibility\cite{ALICE:2023ama}.}
     \label{fig:cross_section_pp_ch}
\end{figure*}

\begin{figure*}
     \centering
     \includegraphics[width=0.49\textwidth]{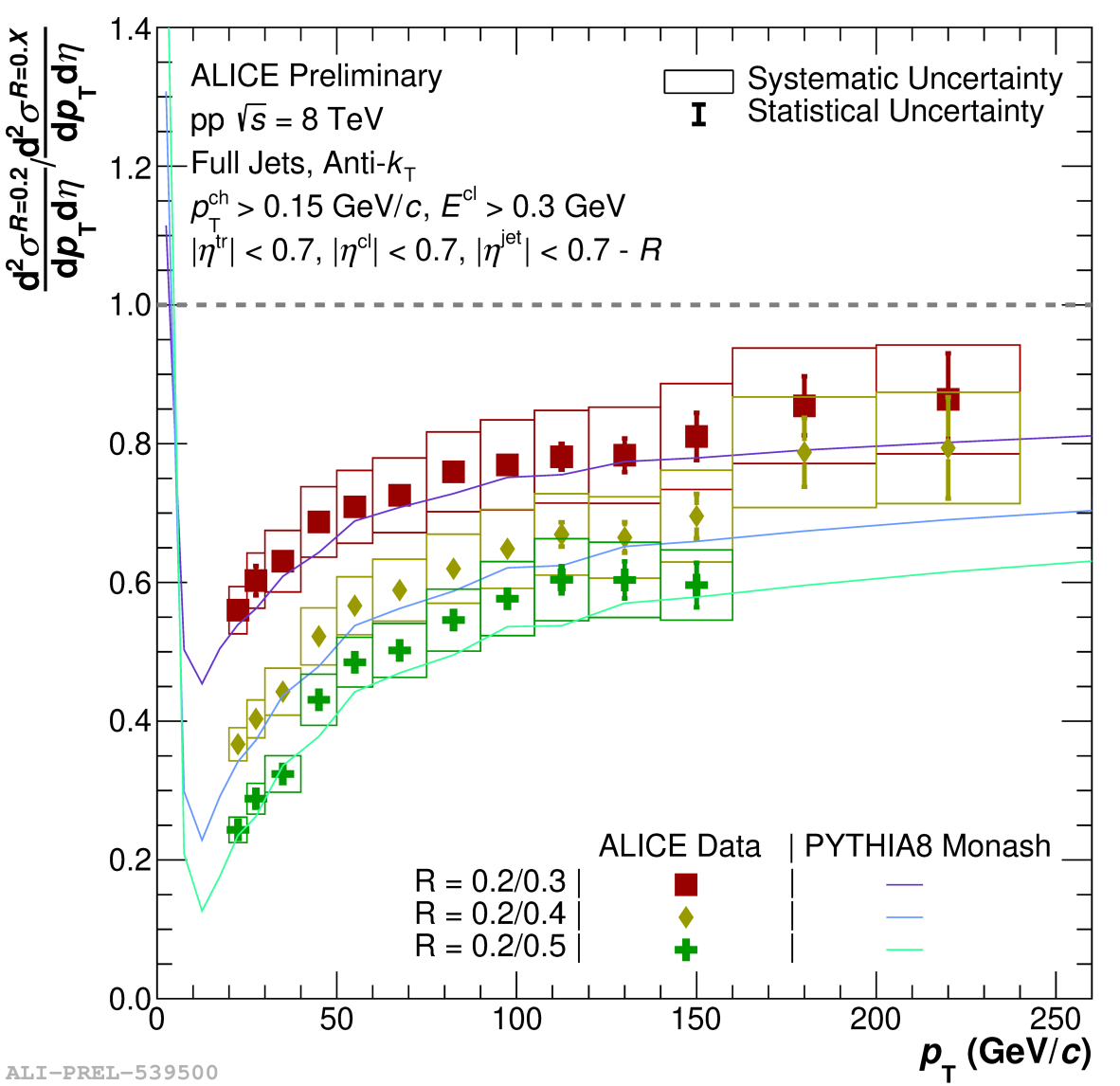}
     \includegraphics[width=0.49\textwidth]{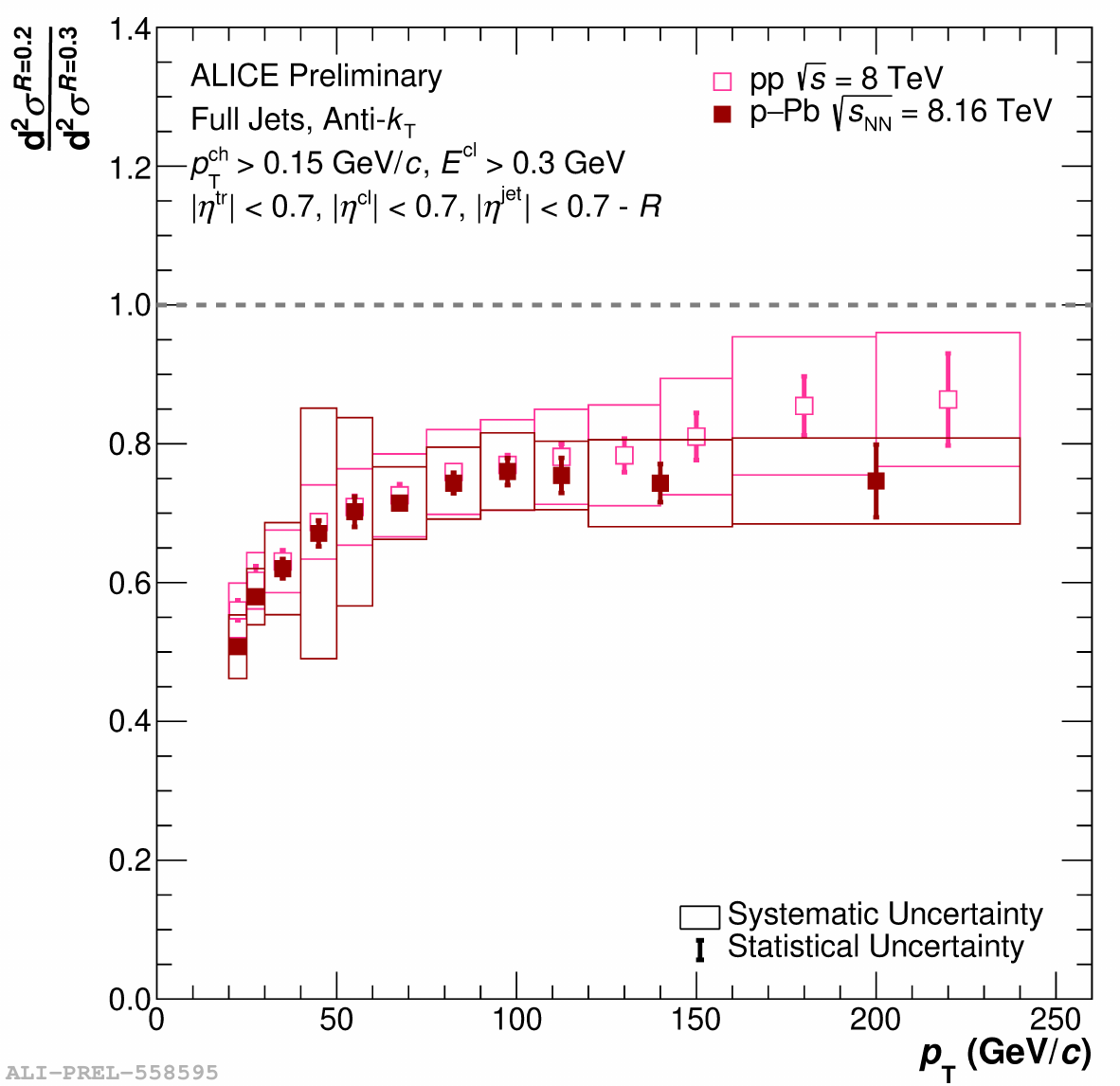}
     \caption{Cross-section ratios using full jets in pp collisions at $\sqrt{\it{s}}$ = 8 TeV compared to PYTHIA (left) and compared to full jets in p--Pb collisions at $\sqrt{\it{s}_{\rm NN}}$ = 8.16 TeV (right).}
     \label{fig:cross_section_ratios}
\end{figure*}

Figure~\ref{fig:cross_section_pp_ch} shows the charged jet cross-section in pp collisions as a function of jet transverse momentum at $\sqrt{\it{s}}$ = 5.02 TeV for several jet radii\cite{ALICE:2023ama} compared to Monte Carlo predictions. These include two from JETSCAPE\cite{Putschke:2019yrg} using a leading-order (LO) pQCD calculation, and one from POWHEG\cite{Oleari:2010nx} predictions using a next-to-leading-order (NLO) pQCD calculation. JETSCAPE overpredicts the data, particularly at low momentum, while the POWHEG agrees within uncertainties. The same is true for full jet cross-sections in pp and charged jet cross-sections in p--Pb.

Cross-section ratios allow for error cancellation and are more sensitive to fragmentation and hadronization. These ratios, taken for R = 0.2 jets to larger jet radii, are depicted in Figure~\ref{fig:cross_section_ratios} for pp collisions (R = 0.2 to R = 0.3-0.5) with comparisons to PYTHIA8\cite{SJOSTRAND2015159} on the left and full jets in p-Pb collisions on the right. Uncertainties for full jets are primarily influenced by tracking efficiency and unfolding. The results exhibit good agreement with predictions from both LO and NLO\cite{ALICE:2023ama}, showcasing consistent behavior and fragmentation patterns across various systems. As momentum increases, ratios approach unity, indicating predominant jet momentum within an R = 0.2 cone, while at lower momentum, the ratios decline, signifying less collimated jets.

\begin{figure}[h]
     \centering
     \includegraphics[width=0.6\textwidth]{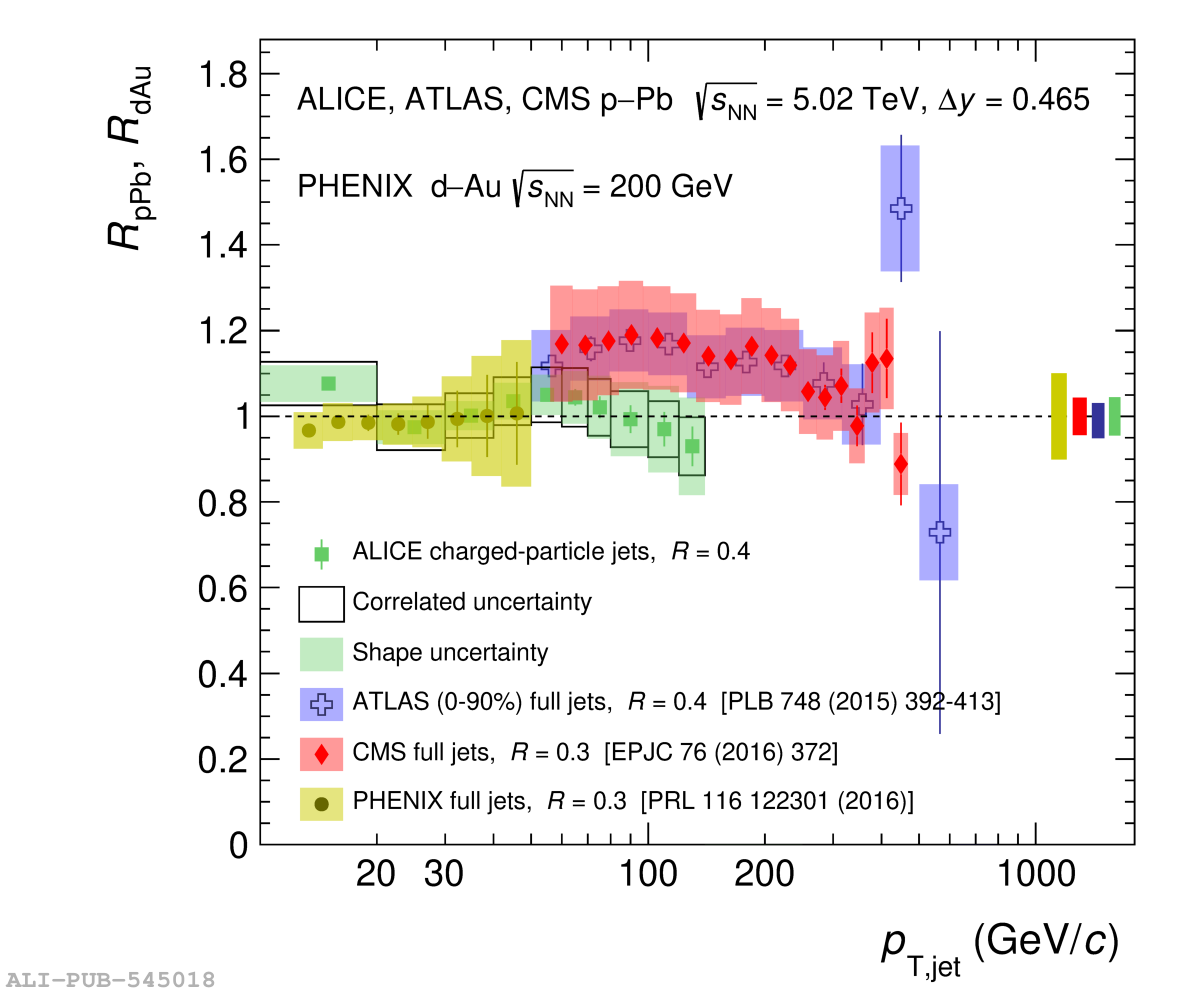}
     \caption{Comparison of the nuclear modification factors of jets in p--Pb at $\sqrt{\it{s}_{\rm NN}}$ = 5.02 TeV and d--Au at $\sqrt{\it{s}_{\rm NN}}$ = 200 GeV at the LHC
     and RHIC, respectively\cite{ALICE:2023ama}.}
     \label{fig:RpPb_comparison}
\end{figure}

The $R_{\rm pPb}$ is the ratio of the cross-section in p--Pb to pp, scaled by the number of binary collisions. Deviation from unity suggests CNM effects or jet quenching. For charged jets at $\sqrt{\it{s}{\rm NN}}$ = 5.02 TeV with jet radii $R$ = 0.2--0.4, $R{\rm pPb}$ remains consistent with unity within uncertainties, indicating no modification in p--Pb collisions. This aligns with CMS\cite{CMS:2016svx} and ATLAS\cite{ATLAS:2014cpa} findings at 5.02 TeV, shown in Figure~\ref{fig:RpPb_comparison}, along with a PHENIX\cite{PHENIX:2015fgy} measurement at 200 GeV. ALICE extends results to a jet {\it{p}}\textsubscript{T} of 10 GeV/c, complementing ATLAS and CMS measurements. ALICE uses charged jets, while ATLAS and CMS use full jets, and their measurements employ an extrapolated pp reference from $\sqrt{\it{s}}$ = 7 TeV.

\section{Conclusions}
\label{conclusions}

Jet cross-section measurements are crucial for comprehending overall jet formation and constraining theoretical models. Charged and full jet measurements, with distinct strengths, complement each other. Cross-section ratios are adequately described by LO pQCD, but NLO is necessary for standalone cross-section descriptions, as outlined in~\cite{Soyez:2011np}. The $R_{\rm pPb}$, consistent with unity within uncertainties, signifies no modification in p–Pb collisions, aligning with prior measurements.

\clearpage

\bibliography{references-dissertation}

\end{document}